\documentclass[aps,pra,showpacs,superscriptaddress,preprint]{revtex4}
\usepackage{graphicx}
\catcode`ð=\active
\defð{\u{g}}
\catcode`Ð=\active
\defÐ{\u{G}}
\catcode`Ý=\active
\defÝ{\. I}
\catcode`ö=\active
\defö{\"{o}}
\catcode`Ö=\active
\defÖ{\"O}
\catcode`ü=\active
\defü{\"{u}}
\catcode`Ü=\active
\defÜ{\"{U}}
\catcode`Þ=\active
\defÞ{\c{S}}
 \catcode`þ=\active
\defþ{\c{s}}
\catcode`ý=\active
\defý{{\i}}
\catcode`ç=\active
\defç{\d{c}}
\catcode`Ç=\active
\defÇ{\d{C}}

\begin{document}
\title{Approximate Analytical Solutions of the Effective Mass Dirac Equation
for the generalized Hulth\'{e}n Potential with any $\kappa$-Value}
\author{\small Altuð Arda}
\email[E-mail: ]{arda@hacettepe.edu.tr}\affiliation{Department of
Physics Education, Hacettepe University, 06800, Ankara,Turkey}
\author{\small Ramazan Sever}
\email[E-mail: ]{sever@metu.edu.tr}\affiliation{Department of
Physics, Middle East Technical  University, 06800, Ankara,Turkey}
\author{\small Cevdet Tezcan}
\email[E-mail: ]{ctezcan@baskent.edu.tr}\affiliation{Faculty of
Engineering, Baþkent University, Baglýca Campus, Ankara,Turkey}
\begin{abstract}
The Dirac equation, with position-dependent mass, is solved
approximately for the generalized Hulth\'{e}n potential with any
spin-orbit quantum number $\kappa$. Solutions are obtained by
using an appropriate coordinate transformation, reducing the
effective mass Dirac equation to a Schr\"{o}dinger-like
differential equation.   The Nikiforov-Uvarov method is used in
the calculations to obtain energy eigenvalues and the
corresponding wave functions. Numerical results are compared with
those given in the literature. Analytical results are also
obtained for the case of constant mass and the results are in good
agreement with the literature.
\end{abstract}

\maketitle

\newpage

\section{Introduction}
In recent years, following work done by von Roos, and Levy-Leblond
[1, 2] on relativistic and non-relativistic motion, the
position-dependent mass (PDM) formalism has recieved attention in
quantum physics. The solution of the Schr\"{o}dinger equation with
effective mass is a useful starting-point for the investigation of
some physical systems, such as the electric properties of quantum
wells and quantum dots [3],  impurities in crystals [4-6], and
semiconductor heterostructures [7]. In general, for both
relativistic and non-relativistic cases, the energy spectra and
corresponding eigenfunctions have been studied by many authors
using different methods and potentials: deformed algebras applied
to a Coulomb problem [8], analysis within supersymmetric quantum
mechanics [9-11], point canonical transformation to study
different classes of potentials [12], non-relativistic Green's
functions applied to the harmonic oscillator [13], the Dirac
equation in the Coulomb field [14], hyperbolic-type potentials
[15], the Morse potential[16], and the Scarf II potential [17].

Another interesting area that has received a lot of attention is
the investigation of solutions to the Dirac equation when the
Dirac Hamiltonian has either spin or pseudospin symmetry [18]. The
Dirac Hamiltonian of a particle with mass $\mu$ moving in scalar,
$V_s(r)$, and vector, $V_v(r)$, potentials is invariant for two
cases: the sum or the difference of the potentials is nearly zero.
Important features of these symmetries provide an understanding of
the structure of deformed nuclei, including deformation and
superdeformation [19, 20], and enable construction of an effective
shell-model coupling scheme [21-28].

Recently, pseudospin and spin symmetry have been studied for
various potentials, such as the Morse potential [29-31], the
Woods-Saxon potential [32], the Coulomb potential [33], and the
harmonic potential [34-36]. Some authors have also solved the
Dirac equation in the context of  pseudospin symmetry under the
effect of the Eckart potential [37, 38], and the Pöschl-Teller
potential [39, 40], for  spin-orbit quantum number $\kappa=1$
and/or any $\kappa$-value. In Ref. [41], the bound states of both
the Klein-Gordon and Dirac equations for the Hulth\'{e}n potential
are studied by writing these equations as a Riemann-type equation.
In Ref. [42], the energy spectra of the Dirac equation for equal
scalar and vector parts of the Hulth\'{e}n potential is studied
using a perturbative approach. In addition, the Dirac equation is
solved for the Hulth\'{e}n potential, for both spin and pseudospin
cases, using the asymptotic iteration method [43].

In the present work, we solve the effective mass Dirac equation
for the generalized Hulth\'{e}n potential within the framework of
an approximation to the $\kappa(\kappa-1)/r^2$ term. We give the
energy eigenvalue equation and the corresponding eigenfunctions
for any spin-orbit quantum number $\kappa$. We also obtain the
energy eigenvalue equation in the case of constant mass. We give
separately the Dirac solutions for the cases where spin and
pseudospin symmetries are taken into account. We apply the
parametric generalization of the Nikiforov-Uvarov (NU) method to
obtain the energy eigenvalue equations and the eigenfunctions of
the generalized Hulth\'{e}n potential. The NU method is a powerful
method for solving second order differential equations [44], like
methods based on Lie algebras [45]. In Ref. [46], this approach
was generalized to a theory based on spectrum generating algebras
including relativistic effects. The NU method can be used, for
both non-relativistic and  relativistic cases, as a tool to find
the energy spectrum and corresponding wave functions.

\section{Dirac Equation}
The Dirac equation for a spin-$\frac{1}{2}$ particle with mass
$\mu$ moving in scalar $V_s(r)$, and vector $V_v(r)$ potentials is
written as

\begin{eqnarray}
[\alpha.\textbf{P}+\beta[\mu+V_s(r)]+V_v(r)-E]\Psi(r)=0,
\end{eqnarray}
where $E$ is the relativistic energy of the particle, $\textbf{P}$
is three-momentum, and $\alpha$ and $\beta$ are $4 \times 4$ Dirac
matrices [47] defined, respectively as,

\begin{eqnarray}
\alpha=\Bigg(\begin{array}{cc}
0 & \sigma \\
\sigma & 0
\end{array}\Bigg),\,\,\,\,\,\,\,
\beta=\Bigg(\begin{array}{cc}
0 & I \\
-I & 0
\end{array}\Bigg)
\end{eqnarray}

Here, $\sigma$ is a three-vector whose components are Pauli
matrices, and $I$ denotes the $2 \times 2$ unit matrix. For a
spherically symmetric potential, the total angular momentum
$\textbf{J}$, and spin-orbit operator
$\hat{K}=-\beta(\sigma.\textbf{L}+1)$ commute with the Dirac
Hamiltonian, where $\textbf{L}$ is the orbital angular momentum
operator. The eigenvalues of the operator $\hat{K}$ are
$\kappa=\pm(j+1/2)$; $\kappa=-(j+1/2)<0$ stands for the aligned
spin $j=\ell+1/2$, and $\kappa=(j+1/2)>0$ stands for the unaligned
spin $j=\ell-1/2$. On the other hand, the total angular quantum
number can be written in terms of the pseudo-orbital angular
momentum $\tilde{\ell}=\ell+1$ and the pseudospin angular momentum
$\tilde{s}=1/2$ as $j=\tilde{\ell}+\tilde{s}$. The spin-orbit
quantum number $\kappa$ is related to the orbital angular quantum
number by the expressions $\kappa(\kappa+1)=\ell(\ell+1)$, and
$\kappa(\kappa-1)=\tilde{\ell}(\tilde{\ell}+1)$ for a given
$\kappa=\pm1, \pm2, \ldots$. The spherically symmetric Dirac
spinor can be written in terms of upper and lower components,

\begin{eqnarray}
\Psi_{n \kappa}(r)=\,\frac{1}{r}\,\Bigg(\begin{array}{c} \,F_{n
\kappa}\,(r)Y_{jm}^{\ell}(\theta,\phi) \\
iG_{n \kappa}\,(r)Y_{jm}^{\tilde{\ell}}(\theta,\phi)
\end{array}\Bigg)\,,
\end{eqnarray}
where $Y_{jm}^{\ell}(\theta,\phi)$, and
$Y_{jm}^{\tilde{\ell}}(\theta,\phi)$ are  spherical harmonics, and
$F_{n \kappa}\,(r)/r$, and $G_{n \kappa}\,(r)/r$ are the radial
parts of the upper and lower components. Substituting Eq. (3) into
Eq. (1) one can write the Dirac equation as a set of two coupled
differential equations in terms of $F_{n \kappa}\,(r)$ and $G_{n
\kappa}\,(r)$

\begin{eqnarray}
\Bigg(\,\frac{d}{dr}\,+\,\frac{\kappa}{r}\,\Bigg)\,F_{n
\kappa}\,(r)&=&[\mu+E_{n \kappa}-V_{-}(r)]G_{n \kappa}\,(r)\,,\\
\Bigg(\,\frac{d}{dr}\,-\,\frac{\kappa}{r}\,\Bigg)\,G_{n
\kappa}\,(r)&=&[\mu-E_{n \kappa}+V_{+}(r)]F_{n \kappa}\,(r)\,,
\end{eqnarray}
where $V_{-}(r)=V_{v}\,(r)-V_s\,(r)$, and
$V_{+}(r)=V_{v}\,(r)+V_s\,(r)$. Using Eq. (4) for the upper
component, and substituting into Eq. (5), we obtain generalized
differential equations for the position-dependent mass case,

\begin{eqnarray}
\Bigg(\,\frac{d^2}{dr^2}-\,\frac{\kappa(\kappa+1)}{r^2}-[\mu+E_{n\kappa}-V_{-}(r)]
[\mu&-&E_{n\kappa}+V_{+}(r)]\nonumber\\&-&\frac{(\frac{d\mu(r)}{dr}-\frac{dV_{-}(r)}{dr})\,(\frac{d}{dr}
\,+\,\frac{\kappa}{r})}
{\mu+E_{n\kappa}-V_{-}(r)}\Bigg)F_{n\kappa}(r)=0\,,
\end{eqnarray}

\begin{eqnarray}
\Bigg(\,\frac{d^2}{dr^2}-\,\frac{\kappa(\kappa-1)}{r^2}-[\mu+E_{n\kappa}-V_{-}(r)]
[\mu&-&E_{n\kappa}+V_{+}(r)]\nonumber\\&-&\frac{(\frac{d\mu(r)}{dr}+\frac{dV_{+}(r)}{dr})\,(\frac{d}{dr}
\,-\,\frac{\kappa}{r})}
{\mu-E_{n\kappa}+V_{+}(r)}\Bigg)G_{n\kappa}(r)=0\,.
\end{eqnarray}
Here, the energy eigenvalues depend on the quantum numbers $n$,
and $\kappa$, and also on the quantum number $\tilde{\ell}$
according to the relation
$\kappa(\kappa-1)=\tilde{\ell}(\tilde{\ell}+1)$. To solve these
equations, we approximate the centrifugal term. Thus, the energy
spectra and the corresponding eigenfunctions can be obtained
analytically by using the NU method.

\section{Parametric Formulation of Nikiforov-Uvarov Method}
We briefly give the mathematical background required for the
parametric NU method [48], where the general form of the
Schr\"{o}dinger-like equation for any potential is written as

\begin{eqnarray}
\left[\frac{d^{2}}{ds^{2}}+\frac{\alpha_{1}-\alpha_{2}s}{s(1-\alpha_{3}s)}
\frac{d}{ds}+\frac{-\xi_{1}s^{2}+\xi_{2}s-\xi_{3}}{[s(1-\alpha_{3}s)]^{2}}\right]\psi(s)=0.
\end{eqnarray}
Comparing Eq. (8) with the general form of the basic equation of
the method, we obtain

\begin{eqnarray}
\tilde{\tau}(s)=\alpha_{1}-\alpha_{2}s\,\,\,\,;\sigma(s)=s(1-\alpha_{3}s)\,\,\,\,;
\tilde{\sigma}(s)=-\xi_{1}s^{2}+\xi_{2}s-\xi_{3}\,.
\end{eqnarray}
Substituting Eq. (9) into the polynomial $\pi(s)=\frac
{\left(\sigma^{\prime}-\tilde{\tau}\right)}{2}\pm
\sqrt{\left(\frac{\sigma^{\prime}-\tilde{\tau}}{2}\right)^{2}-
\tilde{\sigma}+k\sigma}$, we obtain

\begin{eqnarray}
\pi(s)=\alpha_{4}+\alpha_{5}s\pm\sqrt{(\alpha_{6}-k\alpha_{3})s^{2}+(\alpha_{7}+k)s+\alpha_{8}}\,,
\end{eqnarray}
where $\alpha_{4}=\frac{1}{2}\,(1-\alpha_{1})$,
$\alpha_{5}=\frac{1}{2}\,(\alpha_{2}-2\alpha_{3})$,
$\alpha_{6}=\alpha_{5}^{2}+\xi_{1}$,
$\alpha_{7}=2\alpha_{4}\alpha_{5}-\xi_{2}$, and
$\alpha_{8}=\alpha_{4}^{2}+\xi_{3}$.

In the NU-method, the function under the square root in Eq. (10)
must be the square of a polynomial [48]. This condition gives the
roots of the parameter $k$,

\begin{eqnarray}
k_{1,2}=-(\alpha_{7}+2\alpha_{3}\alpha_{8})\pm2\sqrt{\alpha_{8}\alpha_{9}}\,,
\end{eqnarray}
with
$\alpha_{9}=\alpha_{3}\alpha_{7}+\alpha_{3}^{2}\alpha_{8}+\alpha_{6}$.
For
$k=-(\alpha_{7}+2\alpha_{3}\alpha_{8})-2\sqrt{\alpha_{8}\alpha_{9}}$,
$\pi(s)$ becomes

\begin{eqnarray}
\pi(s)=\alpha_{4}+\alpha_{5}s-\left[(\sqrt{\alpha_{9}}+\alpha_{3}\sqrt{\alpha_{8}}\,)s-\sqrt{\alpha_{8}}\,\right]\,,
\end{eqnarray}
and also

\begin{eqnarray}
\tau(s)=\alpha_{1}+2\alpha_{4}-(\alpha_{2}-2\alpha_{5})s-2\left[(\sqrt{\alpha_{9}}
+\alpha_{3}\sqrt{\alpha_{8}}\,)s-\sqrt{\alpha_{8}}\,\right].
\end{eqnarray}

To satisfy the condition that the derivative of the function
$\tau(s)$ should be negative, we impose

\begin{eqnarray}
\tau^{\prime}(s)&=&-(\alpha_{2}-2\alpha_{5})-2(\sqrt{\alpha_{9}}+\alpha_{3}\sqrt{\alpha_{8}}\,)
\nonumber \\
&=&-2\alpha_{3}-2(\sqrt{\alpha_{9}}+\alpha_{3}\sqrt{\alpha_{8}}\,)\quad<0.
\end{eqnarray}

The energy eigenvalue equation is written as [48]

\begin{eqnarray}
\alpha_{2}n-(2n+1)\alpha_{5}&+&(2n+1)(\sqrt{\alpha_{9}}+\alpha_{3}\sqrt{\alpha_{8}}\,)+n(n-1)\alpha_{3}\nonumber\\
&+&\alpha_{7}+2\alpha_{3}\alpha_{8}+2\sqrt{\alpha_{8}\alpha_{9}}=0.
\end{eqnarray}
and the equality
$\sigma'(s)\rho(s)+\sigma(s)\rho'(s)=\tau(s)\rho(s)$ gives

\begin{eqnarray}
\rho(s)=s^{\alpha_{10}-1}(1-\alpha_{3}s)^{\frac{\alpha_{11}}{\alpha_{3}}-\alpha_{10}-1}\,.
\end{eqnarray}
This equation together with
$y_{n}(s)\sim\frac{1}{\rho(s)}\frac{d^{n}}{ds^{n}}
\left[\sigma^{n}(s)~\rho(s)\right]$  gives

\begin{eqnarray}
y_{n}(s)=P_{n}^{(\alpha_{10}-1,\frac{\alpha_{11}}{\alpha_{3}}-\alpha_{10}-1)}(1-2\alpha_{3}s)\,,
\end{eqnarray}
where $\alpha_{10}=\alpha_{1}+2\alpha_{4}+2\sqrt{\alpha_{8}\,}$,
$\alpha_{11}=\alpha_{2}-2\alpha_{5}+2(\sqrt{\alpha_{9}}+\alpha_{3}\sqrt{\alpha_{8}})$
and $P_{n}^{(\alpha,\beta)}(1-2\alpha_{3}s)$ are the Jacobi
polynomials. The equality
$\frac{\phi^{\prime}(s)}{\phi(s)}=\frac{\pi(s)}{\sigma(s)}$ [48]
gives

\begin{eqnarray}
\phi(s)=s^{\alpha_{12}}(1-\alpha_{3}s)^{-\alpha_{12}-\frac{\alpha_{13}}{\alpha_{3}}}\,,
\end{eqnarray}
and the general solution $\psi(s)=\phi(s)y(s)$ becomes

\begin{eqnarray}
\psi(s)=s^{\alpha_{12}}(1-\alpha_{3}s)^{-\alpha_{12}-\frac{\alpha_{13}}{\alpha_{3}}}
P_{n}^{(\alpha_{10}-1,\frac{\alpha_{11}}{\alpha_{3}}-\alpha_{10}-1)}(1-2\alpha_{3}s)\,,
\end{eqnarray}
where $\alpha_{12}=\alpha_{4}+\sqrt{\alpha_{8}}$ and
$\alpha_{13}=\alpha_{5}-(\sqrt{\alpha_{9}}+\alpha_{3}\sqrt{\alpha_{8}})\,$
[48].

\section{Bound-State Solutions}
In order to solve the Dirac equation, including the term
proportional to $1/r^2$, we set the vector and scalar potentials
as the generalized Hulth\'{e}n potential [31, 49] of the forms

\begin{eqnarray}
V_{v}(r)&=&V_{0}\frac{e^{-2\alpha r}}{e^{-2\alpha
r}-1}\,\,\,,V_{0}>0\,,\\V_{s}(r)&=&-S_{0}\frac{e^{-2\alpha
r}}{e^{-2\alpha r}-1}\,\,\,,S_{0}>0\,,
\end{eqnarray}
where $\alpha$ is the screening parameter (positive), the constant
parameter $S_0$ denotes the scalar and $V_0$ denotes the vector
part of the potential, respectively. Using the last two equations,
we obtain

\begin{eqnarray}
V_{-}(r)&=&(V_{0}+S_{0})\frac{e^{-2\alpha r}}{e^{-2\alpha r}-1}\,,\\
V_{+}(r)&=&(V_{0}-S_{0})\frac{e^{-2\alpha r}}{e^{-2\alpha r}-1}\,.
\end{eqnarray}
where the ``effective" potential $V_{-}(r)$ is a repulsive one in
the relativistic region [18, 26-28, 35, 50, 51]. It is well known
that the Hulth\'{e}n potential gives bound states in the
non-relativistic region only if the potential parameters satisfy
the condition that $\delta^2<V_0<4\delta^2$, where
$\delta^2=4\alpha^2$, in the absence of the scalar part [41].

We use the equality $d\mu(r)/dr=-\,dV_{+}(r)/dr$ to eliminate the
last term in Eq. (7), and obtain the mass function which can be
written as

\begin{eqnarray}
\mu(r)=\mu_0-\frac{\mu_1}{1-e^{2\alpha r}}\,.
\end{eqnarray}

The mass function has the same form as the Hulth\'{e}n potential,
where $\mu_0$ denotes the integral constant, and
$\mu_{1}=V_{0}-S_{0}$. Thus, the parameter $\mu_{1}$ contains
contributions coming from the scalar, as well as the vector part
of the potential. The parameter $\mu_{0}$ corresponds to the rest
mass of the Dirac particle. Substituting Eqs. (22), (23) and (24)
into Eq. (7), we get

\begin{eqnarray}
\Big\{\,\frac{d^2}{dr^2}\,-\,\frac{\kappa(\kappa-1)}{r^2}\,&-&\Big(\mu_{0}-\mu_{1}\frac{e^{-2\alpha
r}}{e^{-2\alpha r}-1}-(V_{0}+S_{0})\frac{e^{-2\alpha
r}}{e^{-2\alpha r}-1}+E_{nk}\Big)\nonumber
\\&\times&\Big( \mu_0-\mu_{1}\frac{e^{-2\alpha
r}}{e^{-2\alpha r}-1}+(V_0-S_{0})\frac{e^{-2\alpha r
}}{e^{-2\alpha r}-1}-E_{nk}\Big)\Big\}G_{n\kappa}(r)=0\,,
\end{eqnarray}

The following approximation is used for $1/r^2$ term [31, 52]
\begin{eqnarray}
\frac{1}{r^2}\approx\,\frac{4\alpha^2 e^{-2\alpha
r}}{(1-e^{-2\alpha r})^2}\,,
\end{eqnarray}
We compute $1/r^2$ expanding into series. This will provide a
physical result for $\alpha=0$.

Defining a new variable $s=e^{-2\alpha r}$, we have

\begin{eqnarray}
\Bigg\{\,\frac{d^2}{ds^2}&+&\frac{1-s}{s(1-s)}\frac{d}{ds}+\frac{1}{[s(1-s)]^2}
\Big[\eta^2(E^{2}_{nk}-\mu^{2}_{0})\nonumber\\&-&
2\eta^2[\mu_{0}(S_{0}-\mu_{0})-E_{n\kappa}V_{0}+E^{2}_{nk}+2\alpha^2\kappa(\kappa-1)+\mu_{1}\mu_{0}]s
\nonumber\\&-&\eta^2[\mu^{2}_{0}-E^{2}_{nk}+2E_{n\kappa}V_{0}-2S_{0}\mu_{0}+\mu_{1}(\mu_{1}+2S_{0}-2\mu_{0})
\nonumber\\&-&V^2_{0}+S^2_{0}]s^2\Big] \Bigg\}G_{n\kappa}(s)=0\,,
\end{eqnarray}
where $\eta^2=1/4\alpha^2$\,. Comparing Eq. (27) with Eq. (8), we
get the parameter set given in Section II:

\begin{eqnarray}
\begin{array}{ll}
\alpha_1=1\,, &
-\xi_1=\eta^2[\mu^{2}_{0}-E^{2}_{nk}+2E_{n\kappa}V_{0}-2S_{0}\mu_{0}+\mu_{1}(\mu_{1}+2S_{0}-2\mu_{0})
-V^2_{0}+S^2_{0}] \\
\alpha_2=1\,, &
\xi_2=-2\eta^2[\mu_{0}(S_{0}-\mu_{0})-E_{n\kappa}V_{0}+E^{2}_{nk}+2\alpha^2\kappa(\kappa-1)+\mu_{1}\mu_{0}] \\
\alpha_3=1\,, &
-\xi_3=\eta^2(E^{2}_{nk}-\mu^{2}_{0}) \\ \alpha_4=0\,, & \alpha_5=-\,\frac{1}{2} \\
\alpha_6=\xi_1+\frac{1}{4}\,, & \alpha_7=-\xi_2 \\
\alpha_8=\xi_3\,, & \alpha_9=\xi_1-\xi_2+\xi_3+\frac{1}{4} \\
\alpha_{10}=1+2\sqrt{\xi_3}\,, & \alpha_{11}=2+2(\,\sqrt{\xi_1-\xi_2+\xi_3+\frac{1}{4}\,}+\sqrt{\xi_3}\,) \\
\alpha_{12}=\sqrt{\xi_3}\,, &
\alpha_{13}=-\frac{1}{2}-(\,\sqrt{\xi_1-\xi_2+\xi_3+\frac{1}{4}\,}+\sqrt{\xi_3}\,)
\end{array}
\end{eqnarray}

We can easily obtain the energy eigenvalue equation of the
generalized Hulth\'{e}n potential for any $\kappa$ value from Eq.
(15)

\begin{eqnarray}
2\eta\sqrt{\mu^2_{0}-E^2_{n\kappa}\,}&=&\frac{\eta^2}{N}\Big(2E_{n\kappa}V_{0}+\mu_{1}(\mu_{1}+2S_{0}-2\mu_{0})
-V^2_{0}+S^2_{0}-2\mu_{0}S_{0}\Big)-N\,,\nonumber\\
\end{eqnarray}
where

\begin{eqnarray}
N=\frac{1}{2}(2n+1)+\sqrt{\eta^2[\mu_{1}(\mu_{1}+2S_{0})-V^2_{0}+S^2_{0}
+4\alpha^2\kappa(\kappa-1)]+\frac{1}{4}\,}\,.
\end{eqnarray}

The numerical results for different quantum numbers $(n,\kappa)$
are listed in Table I. We list the eigenvalues $E_{n\kappa}$ for
constant mass ($\mu_{1}=0$) and for two different values of
$\mu_{1}$ to see the effect of the spatially dependent mass. The
results given for constant mass are compared with results reported
in the literature. For the constant mass case, we see good
agreement with the results given in Ref. [42], where the
relativistic energy $W$ is stated as $W=E+m$.

The lower spinor component can be obtained from Eq. (19)

\begin{eqnarray}
G_{n\kappa}(s)=s^{\epsilon_{n\kappa}}(1-s)^{1/2+\delta}P^{(2\epsilon,\,2\delta)}_{n}(1-2s)\,,
\end{eqnarray}
where
$\epsilon_{n\kappa}=\sqrt{\eta^2(\mu_{0}-E^2_{\kappa})\,}$\,, and
$\delta=\sqrt{\eta^2[\mu_{1}(\mu_{1}+2S_{0})-V^2_{0}+S^2_{0}
+4\alpha^2\kappa(\kappa-1)]+\frac{1}{4}\,}$\,.

From the last equation, and using Eq. (5), the upper spinor
component can be expressed as

\begin{eqnarray}
F_{n\kappa}(s)&=&\frac{2\alpha
s^{\epsilon_{n\kappa}}(1-s)^{\frac{1}{2}+\delta}}{\mu(s)-E_{n\kappa}+\Sigma(s)}
\Big\{\Big[\frac{s}{1-s}\,(\frac{1}{2}+\delta)-\epsilon_{n\kappa}+\frac{\kappa}{lns}\Big]P_{n}^{(2\epsilon_{n\kappa}\,,
\,2\delta)}(1-2s)\nonumber\\&-&\frac{1}{2}\,(n+2\epsilon_{n\kappa}+2\delta+1)P_{n-1}^{(1+2\epsilon_{n\kappa}\,,
\,1+2\delta)}(1-2s)\Big\}\,.
\end{eqnarray}

We consider the case of constant mass to discuss the compatibility
of our results. Setting $\mu_1=0$ in Eq. (24) while keeping in
mind that $V_{+}(r)=C=const.$ for pseudospin symmetry, and
following the same procedure ($s=e^{-2\alpha r}$), we obtain the
energy eigenvalue equation from Eq. (7)

\begin{eqnarray}
(\mu_{0}+E_{n\kappa})(\mu_{0}-E_{n\kappa}+C)
&=&\frac{1}{4\eta^2}\bigg[\frac{1}{2}\,(2n+1)+\sqrt{\kappa(\kappa-1)+\frac{1}{4}\,}\nonumber\\
&+&\eta^2V\frac{\mu_{0}-E_{n\kappa}+C}{\frac{1}{2}\,(2n+1)+\sqrt{\kappa(\kappa-1)+\frac{1}{4}\,}}\bigg]^2\,,
\end{eqnarray}
where $V=V_{0}+S_{0}$\,. The result in the case of constant mass
is  Eq. (47) of Ref. [43] for $S_0 \rightarrow 0$\, under the
exact pseudospin symmetry.

We now briefly give the bound state solutions for the constant
mass case under spin symmetry. Setting $V_{-}(r)=C=const.$, and
$\mu_{1}=0$, we get the bound state solutions under the exact spin
symmetry for the case of constant mass from Eq. (6) as
($s=e^{-2\alpha r}$)

\begin{eqnarray}
(\mu_{0}-E_{n\kappa})(\mu_{0}+E_{n\kappa}-C)
&=&\frac{1}{4\eta^2}\bigg[\frac{1}{2}\,(2n+1)+\sqrt{\kappa(\kappa+1)+\frac{1}{4}\,}\nonumber\\
&+&\eta^2V'\frac{\mu_{0}+E_{n\kappa}-C}{\frac{1}{2}\,(2n+1)+\sqrt{\kappa(\kappa+1)+\frac{1}{4}\,}}\bigg]^2\,,
\end{eqnarray}
where $V'=S_{0}-V_{0}$\,.

\section{Conclusion}
We approximately solved the Dirac equation, with
position-dependent mass, for the generalized Hulth\'{e}n potential
with arbitrary spin-orbit quantum number. We  found the eigenvalue
equation, and corresponding two-component spinors in terms of
Jacobi polynomials by using the parametric generalization of the
NU-method within the framework of an approximation to the
$\kappa(\kappa-1)/r^2$ term. We compared the numerical results
with those obtained in the literature and given in Table I. We
showed results for the case of constant mass, and  summarized the
results for two different position-dependent mass values, obtained
when $\mu_{1}=0.005$ and $\mu_{1}=0.0001$. We also obtained the
energy eigenvalue equation for the constant mass case with spin
and pseudospin symmetries, separately. These analytical results
are in agreement with results reported in the literature.

\section{Acknowledgments}
This research was partially supported by the Scientific and
Technical Research Council of Turkey.

\begin{table}
\begin{ruledtabular}
\caption{Energy eigenvalues for different values of $n$ and
$\kappa$\,($\mu_{0}=V_{0}=1$).}
\begin{tabular}{llcccc}
$\mu_{1}$ & $\alpha$ & $n\,\,\,\kappa$ & state & $E_{n\kappa}<0$ & Ref.[47]  \\
\hline
0 & 0.025 & 1\,\,-1 & $1s_{1/2}$ & 0.998068 & -1.993900 \\
& 0.01 &  &  & 0.999691 & -1.999000 \\
& 0.05 &  &  & 0.992258 & -1.975800 \\
& 0.1 &  &  & 0.968772 & -1.905700 \\ \hline
0.005& 0.025 & 1\,\,-1 & $1s_{1/2}$ & --- & --- \\
&  & 1\,\,-2 & $1p_{3/2}$ & 0.998068 & --- \\
&  & 1\,\,-3 & $1d_{5/2}$ & 0.995895 & --- \\
&  & 1\,\,-4 & $1f_{7/2}$ & 0.993424 & --- \\ \hline
& 0.05 & 1\,\,-1 & $1s_{1/2}$ & 0.994106 & --- \\
&  & 1\,\,-2 & $1p_{3/2}$ & 0.987615 & --- \\
&  & 1\,\,-3 & $1d_{5/2}$ & 0.979641 & --- \\
&  & 1\,\,-4 & $1f_{7/2}$ & 0.969946 & --- \\ \hline
& 0.1 & 1\,\,-1 & $1s_{1/2}$ & 0.970550 & --- \\
&  & 1\,\,-2 & $1p_{3/2}$ & 0.945421 & --- \\
&  & 1\,\,-3 & $1d_{5/2}$ & 0.912747 & --- \\
&  & 1\,\,-4 & $1f_{7/2}$ & 0.871793 & --- \\ \hline
0.0001& 0.025 & 1\,\,-1 & $1s_{1/2}$ & 0.998103 & --- \\
&  & 1\,\,-2 & $1p_{3/2}$ & 0.996592 & --- \\
&  & 1\,\,-3 & $1d_{5/2}$ & 0.994653 & --- \\
&  & 1\,\,-4 & $1f_{7/2}$ & 0.992281 & --- \\ \hline
& 0.05 & 1\,\,-1 & $1s_{1/2}$ & 0.992293 & --- \\
&  & 1\,\,-2 & $1p_{3/2}$ & 0.986235 & --- \\
&  & 1\,\,-3 & $1d_{5/2}$ & 0.978413 & --- \\
&  & 1\,\,-4 & $1f_{7/2}$ & 0.968795 & --- \\ \hline
& 0.1 & 1\,\,-1 & $1s_{1/2}$ & 0.968807 & --- \\
&  & 1\,\,-2 & $1p_{3/2}$ & 0.944015 & --- \\
&  & 1\,\,-3 & $1d_{5/2}$ & 0.911464 & --- \\
&  & 1\,\,-4 & $1f_{7/2}$ & 0.870558 & --- \\
\end{tabular}
\end{ruledtabular}
\end{table}

\end{document}